\documentclass[aps,prl,twocolumn,showpacs]{revtex4}

\usepackage{amsmath,epsfig,amsfonts,epsfig,graphicx}

\begin{document}

\title{Discrete Breathers in One-Dimensional Diatomic Granular Crystals}

\author{N. Boechler$^{1}$, G. Theocharis$^{2,1}$, S. Job$^{3,1}$, P. G. Kevrekidis$^{2}$, M. A. Porter$^{4}$ and C. Daraio$^{1}$}

\affiliation{ 
$^1$ Graduate Aerospace Laboratories (GALCIT), California Institute of Technology, Pasadena, CA 91125, USA \\
$^2$ Department of Mathematics and Statistics, University of Massachusetts, Amherst MA 01003-4515, USA \\
$^3$ Supmeca, 3 rue Fernand Hainaut, 93407 Saint-Ouen, France \\
$^4$ Mathematical Institute, University of Oxford, OX1 3LB, UK
}


\begin{abstract}
We report the experimental observation of discrete breathers in a one-dimensional diatomic granular crystal composed of compressed elastic beads that interact via Hertzian contact. We first characterize their effective linear spectrum both theoretically and experimentally. We then illustrate theoretically and numerically the modulational instability of the lower edge of the optical band. This leads to the dynamical formation of long-lived breather structures, whose families of solutions we compute throughout the linear spectral gap. Finally, we observe experimentally such localized breathing modes with quantitative characteristics that agree with our numerical results.
\end{abstract}


\maketitle


{\it Introduction}. Intrinsic localized modes (ILMs), or discrete breathers (DBs), have been a central theme in numerous theoretical and experimental investigations during the past two decades~\cite{camp04,Flach2007}. Their original theoretical proposal in settings such as anharmonic nonlinear lattices \cite{sietak} and the rigorous proof of their existence under fairly general conditions~\cite{macaub} motivated studies of such modes in a diverse host of applications, including charge-transfer solids~\cite{Swanson1999}, antiferromagnets~\cite{Schwarz1999}, superconducting Josephson junctions~\cite{orlando1}, photonic crystals~\cite{photon}, biopolymers~\cite{Xiepeyrard}, micromechanical cantilever arrays~\cite{sievers}, Bose-Einstein condensates~\cite{morsch1}, and more.

Granular crystals, consisting of closely packed ensembles of elastically interacting particles have also recently drawn considerable attention. This broad interest has arisen from their nonlinear contact dynamics and the tunability of their dynamic response to encompass linear, weakly nonlinear, and strongly nonlinear regimes~\cite{nesterenko1,Job2005}. Such flexibility makes them ideal not only as toy models for probing the physics of granular materials but also for the implementation of many engineering applications, including shock and energy absorbing layers~\cite{dar06}, actuating devices~\cite{dev08}, and sound scramblers~\cite{dar05}. Only recently have nonlinear localized modes begun to be explored in granular crystals. Previous studies have focused on metastable breathers in acoustic vacuum~\cite{mohan}, the observation of localized oscillations near a defect~\cite{Job2009,Theo09}, and one-dimensional (1D) diatomic crystals restricted to linear dynamics due to welded sphere contacts~\cite{Hladey}. Understanding and controlling localization in granular systems might lead to new energy harvesting/conversion devices and acoustic filters. 

In this Letter, we investigate the existence, stability, and dynamics of DBs in a compressed 1D diatomic granular crystal using experiments, theory, and numerical simulations. We first detail our experimental setup and theoretical model. We then analyze the system's dynamics in the linear regime, show how a modulational instability generates DBs in the weakly nonlinear regime, and finally provide experimental evidence of their existence.


{\it Experimental setup}. We assemble a 1D diatomic granular crystal by alternating aluminum spheres (Acraball, $6061$-T$6$ type, with radius $R_a=9.525$~mm, mass $m_a=9.75$~g, elastic modulus $E_a=73.5$~GPa, and Poisson ratio $\nu_a=0.33$) and stainless steel spheres (McMaster-Carr, $316$ type, $R_b=R_a$, $m_b=28.84$~g, $E_b=193$~GPa, and $\nu_b=0.3$). The reported values of $E_{a,b}$ and $\nu_{a,b}$ are standard specifications~\cite{ElasticProperties}; we discuss the precise characterization of the effective elastic properties of our system below. We hold the spheres in place using four polycarbonate restraining bars and guide plates. At one end of the crystal, we apply a precompressive force using a lever-mass system. We position an O-$1$ tool steel plate, which we mount on a $1018$ steel angle bracket at the other end of the crystal as a ``wall". We drive dynamical perturbations using a piezoelectric actuator, which we fit on the steel plate. We visualize the evolution of the force-time history of the propagating excitations using periodically-placed calibrated piezo sensors that we embed inside selected particles (preserving the inertia and the bulk stiffness of the original bead~\cite{dar05,Job2005}). We measure the static load using a calibrated strain gauge cell that we place in contact with the lever arm and the last bead of the crystal.


{\it Theoretical model}. We model a 1D diatomic crystal of $N$ spheres as a chain of nonlinear oscillators~\cite{nesterenko1}:
\begin{equation}
	m_{i}\ddot{u}_i=A[\delta_{0}+u_{i-1} - u_{i}]_{+}^{p} - A[\delta_{0}+u_{i} - u_{i+1}]_{+}^{p}\,,
\label{model}
\end{equation}
where $[Y]_+$ denotes the positive part of $Y$, $u_i$ is the displacement of the $i$th sphere (where $i \in \{1,\cdots,N\}$) around the static equilibrium, the masses are $m_{\text{odd}}=m_a$ and $m_{\text{even}}=m_b$, and the coefficient A depend on the exponent $p$ and the geometry/material properties of adjacent beads. The exponent $p=3/2$ yields the Hertz potential law between adjacent spheres~\cite{Johnson1985}. In this case, $A=\left(\frac{3}{4}\frac{1-\nu_a^2}{E_a}+\frac{3}{4}\frac{1-\nu_b^2}{E_b}\right)^{-1}\left(\frac{1}{R_a}+\frac{1}{R_b}\right)^{-1/2}$, and one obtains a static overlap of $\delta_{0}=(F_0/A)^{2/3}$ under a static load $F_0$~\cite{Johnson1985,nesterenko1}. We compute the linear dispersion curve of our system from the linearization of Eq.~\ref{model}. For diatomic crystals, this curve contains two branches (\emph{acoustic} and \emph{optical})~\cite{kittel}. At the edge of the first Brillouin zone---i.e., at $k=\frac{\pi}{2\alpha}$, where $\alpha=R_a+R_b-\delta_0$ is the equilibrium distance between two adjacent beads---the linear spectrum possesses a gap between the upper cutoff $\omega_{1}=\sqrt{2K_2/M}$ of the acoustic branch and the lower cutoff $\omega_{2}=\sqrt{2K_2/m}$ of the optical one; the linear stiffness is $K_2=\frac{3}{2}A^{2/3}F_0^{1/3}$, and we define $M=\max{\{m_a,m_b\}}$ and $m=\min{\{m_a,m_b\}}$. The upper cutoff frequency of the optical band is located at $\omega_{3}=\sqrt{2K_2(1/m+1/M)}$. In Table~\ref{table_th_vs_exp_linear}, we summarize $K_2$, $A$, and the three cutoff frequencies, which we estimate using standard specifications~\cite{ElasticProperties} and compute using a static load of $F_0=20$~N.

\begin{figure}[t]
\begin{center}
\includegraphics{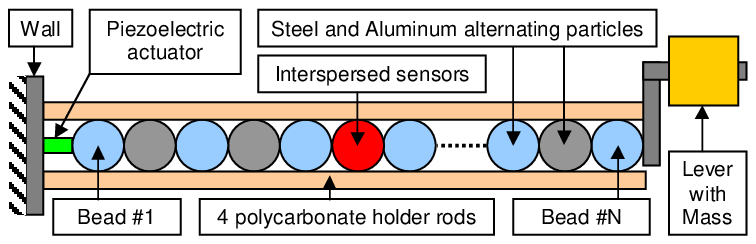}
\includegraphics{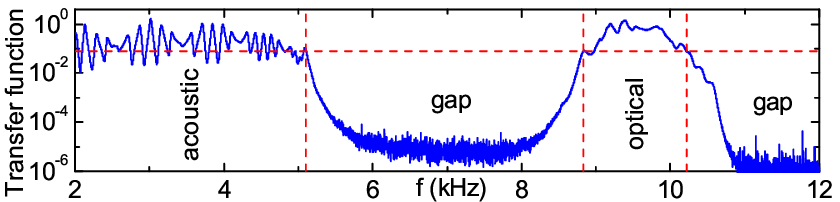}
\end{center}
\caption{\label{setup_spectrum}[Color online] Top panel: Experimental setup. Bottom panel: Experimental phonon spectrum of the 81-bead steel-aluminum diatomic crystal. The horizontal line is half of the low frequency average level and vertical lines indicate the $f_n^{\text{exp}}$ cutoff frequencies given in Table~\ref{table_th_vs_exp_linear}.}
\end{figure}

{\it Linear spectrum}. We experimentally characterize the linear (phonon) spectrum of a diatomic crystal~\cite{DiatPhonon} ($N=81$ and $F_0=20$~N) by applying low-amplitude (peak at approximately $10$~mN), broadband ($2-18$~kHz frequency width) and uniform noise for $800$~ms. We measure the dynamical forces using a sensor located inside the $14$th particle, plus that from the driving voltage and the actuator sensitivity. We then compute the power spectral density (PSD)~\cite{PSD} of the force-sensor, normalize it to the PSD of the driving force, and average the ratio over $8$ acquisitions to obtain the transfer function shown in Fig.~\ref{setup_spectrum}. This spectrum clearly shows forbidden bands (i.e., gaps) and allowed bands bounded by cutoff frequencies. These frequencies match half of the transfer function's low-frequency level, which is determined as the average level in the $2-4$~kHz range. We summarize these frequencies in Table~\ref{table_th_vs_exp_linear}. Matching these frequencies to the theoretical formulas above provides an opportunity to probe the beads' effective parameters $K_2$ and $A$ shown in Table~\ref{table_th_vs_exp_linear} (errorbars indicate the standard deviations from the three frequencies measurements). We find that all of the cutoff frequencies show a systematic upshift of about $9$\% compared to the predictions from standard specifications. We identify four possible explanations for such a systematic bias: (i) the uncertainty in the standard values of material parameters~\cite{ElasticProperties}; (ii) non-Hookean elastic dynamics might lead to slight shift on the nonlinear exponent $p$ and accordingly a large deviation in the coefficient $A$~\cite{Johnson1985}; (iii) imperfect surface smoothness might also induce fluctuations in $p$ and hence $A$~\cite{Coste1999}; and (iv) dissipative mechanisms, such as viscoelasticity and solid friction, can induce stiffening of the interaction potential between particles~\cite{Job2005,Carretero2009}. We also test a shorter ($N=15$) crystal, which showed a higher low-frequency level and lower linear stiffness $K_2\simeq14.80$~N/$\mu$m, in agreement with less dissipation in a shorter crystal.

\begin{table}[t]
\begin{tabular}{|c|c|c|c|c|c|}
\hline       & $f_1$ [kHz] & $f_2$ [kHz] & $f_3$ [kHz] & $K_2$ [N/$\mu$m] & A [N/$\mu$m$^{3/2}$] \\ 
\hline th.   & 4.71        & 8.10        & 9.37        & 12.63            & 5.46                 \\ 
\hline exp.  & 5.11        & 8.83        & 10.22       & $14.95 \pm 0.10$ & $7.04 \pm 0.07$      \\
\hline diff. & +8.5\%      & +9.0\%      & +9.1\%      & +18.4\%          & +28.8\%              \\
\hline 
\end{tabular}
\caption{\label{table_th_vs_exp_linear} Predicted (from standard specifications~\cite{ElasticProperties}) versus measured cutoff frequencies, linear stiffness $K_2$, and coefficient $A$ under a static precompression of $F_0=20$~N.}
\end{table}

\begin{figure}[t]
\begin{center}
\includegraphics[width=8.25cm,height=3.67cm,bb=-188 173 802 617,angle=0,clip]{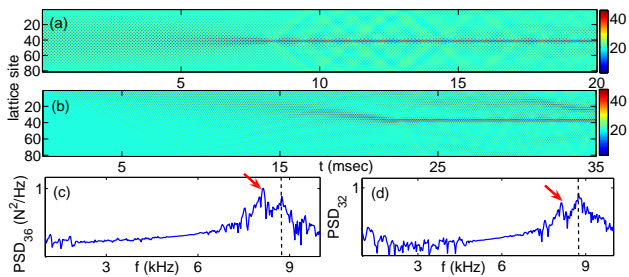}
\end{center}
\caption{\label{MI}[Color online] (a),(b): Spatiotemporal evolution of the forces [N]. (a) Manifestation of the MI of the optical lower cutoff mode. (b) Generation of a DB under conditions relevant to our experimental setup. (c), (d) PSDs of the forces at particles $i=36$ and $i=32$, respectively, for the DB simulation shown in panel (b). Dashed lines indicate the driving frequency $f_{\text{act}}=f_2^{\text{exp}}$, and the arrows indicate the frequency component $f_{b}$ associated with the generated DB.}
\end{figure}

{\it Modulational Instability and DBs}. We now consider the weakly nonlinear dynamics of the granular crystal. If the displacements have small amplitudes relative to those due to precompression, we do a power series expansion of the forces (up to quartic displacement terms) to yield the $K_2-K_3-K_4$ model:
\begin{equation}
	m_{i}\ddot{u}_i=\sum_{k=2}^{4}K_{k}\left[(u_{i+1}-u_{i})^{k-1}-(u_{i}-u_{i-1})^{k-1} \right]\,,
\label{K2K3K4}
\end{equation}
where $K_3=-\frac{3}{8}A^{4/3}F_{0}^{-1/3}$ and $K_4=\frac{3}{48}A^{2}F_{0}^{-1}$. Equation (\ref{K2K3K4}) constitutes a diatomic variant of the Fermi-Pasta-Ulam (FPU) nonlinear oscillator chain~\cite{FPUDiatomic}. Because $\frac{K_3^2}{K_2K_4} > \frac{3}{4}$, the nonlinearity induces an asymmetric localized mode in the gap of the linear spectrum (i.e., one obtains a gap soliton). This arises from the modulational instability (MI) of the optical lower cutoff phonon mode~\cite{Huang1998}. In order to verify this prediction, we solve Eq.~(\ref{model}) numerically using $A_{\text{exp}}$ (see Table~\ref{table_th_vs_exp_linear}) and the optical lower cutoff mode as initial condition. This mode corresponds to the crystal vibration in which the light masses oscillate with frequency $f_{2}^{\text{exp}}$ and the heavy masses are at rest. In order to trigger the instability of this mode, we choose an oscillation amplitude of the light masses that corresponds to an $11.25$~N dynamical peak force. As shown in Fig.~\ref{MI}(a), we observe the MI and the resulting generation of a localized mode with frequency $f_b \simeq 7.95$~kHz at $t \simeq 8$~ms. In order to observe the generation of DBs under conditions relevant to our experimental setup, we also run simulations in which the displacement of the first (actuator-driven) aluminum bead is described by a $30$~ms long square windowed sine with variable oscillation amplitude $B$ and frequency at the optical lower cutoff $f_{\text{act}}=f_{2}^{\text{exp}}$. In Fig.~\ref{MI}(b), we show an example of the spatiotemporal evolution of the forces when $B=0.061\delta_{0}$. In this example, the maximum dynamic force acting on the beads over the first $10$ cycles of the excitation is about $6.5$~N $\simeq 0.325 F_0$. We thus anticipate a weakly nonlinear response that is well described by the $K_2-K_3-K_4$ theory. Initially, a small amplitude excitation is generated and transmitted. During its transmission, the light masses oscillate out of phase, and the heavy ones are practically at rest. At $t\simeq 22 ms$, the excitation experiences MI, which yields a DB which, for these initial conditions, is localized near bead $37$. This nonlinear solution exists even after the actuator is turned off at $t=30$~ms. The PSDs of the force at particles $i=36$ [see Fig.~\ref{MI}(c)] and $i=32$ [see Fig.~\ref{MI}(d)] reveal the presence of a frequency component in the gap at $f_b \simeq 8.14$~kHz $< f_2^{\text{exp}}$. Moreover, it is clear that the dominant frequency is $f_b$ at the center of the localized mode and that $f_{\text{act}}$ dominates $4$ particles away. 

\begin{figure}[t]
\begin{center}
\includegraphics[width=8.25cm,height=3.67cm,angle=0,clip]{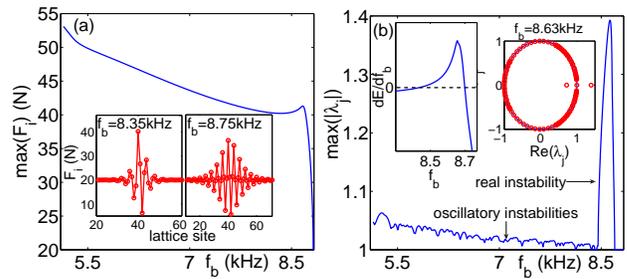}
\end{center}
\caption{\label{MI_actuator}[Color online] Bifurcation diagram of the parameter continuation of the DB solutions. (a) Maximal force of the wave versus frequency $f_b$ (along with profile insets at two values of $f_b$). (b) Maximal deviation of Floquet multipliers from the unit circle, indicating the instability growth strength. The right inset shows a typical multiplier picture, and the left one shows the connection between the strong (real multiplier) instability and the change in sign of $dE/df_b$.}
\end{figure}

{\it Exact solutions and stability of DBs}. We apply Newton's method (see~\cite{Flach2007} and references therein) with free boundary conditions to numerically obtain, with high precision, the above dynamically generated DB waveforms as exact (time-periodic) solutions. We then study their linear stability and frequency dependence (within the spectral gap). Continuing this solution within the gap [i.e., for $f \in (f_{1}^{\text{exp}}, f_{2}^{\text{exp}})$] starting from the optical cutoff mode allows us to trace the entire family of DB solutions. In Fig.~\ref{MI_actuator}(a), we show the maximum force $\max(F_{i})$, which is the experimentally observable parameter of the DB solution, as a function of the DB frequency $f_b$. As $f_b\rightarrow f_2^{\text{exp}}$, $\max(F_{i})\rightarrow F_0$ and the DBs broaden and finally merge with the linear optical lower cutoff mode. In the insets of Fig.~\ref{MI_actuator}(a), we show examples of these solutions with frequencies $f_{b1}=8.35$~kHz and $f_{b2}=8.75$~kHz. To examine the stability of the DB solutions, we compute their Floquet multipliers $\lambda_{j}$~\cite{Flach2007}. If $| \lambda_{j}|=1$ for all $j$, then the DB is linearly stable. In Fig.~\ref{MI_actuator}(b), we show the stability diagram for the family of DB solutions and the corresponding locations of Floquet multipliers in the complex plane for the DB with $f_{b1}=8.83$~kHz. Strictly speaking, the DB is stable only for $f_b \simeq f_2^{\text{exp}}$. Otherwise, the DB family exhibits oscillatory instabilities~\cite{Flach2007,Theo09}. However, the deviations of the unstable eigenvalues from the unit circle are bounded above by $0.08$, and numerical integration of the DBs up to times $100 T$ (where $T$ is their period) reveals their robustness. Importantly, we also find that DB solutions exhibit a strong instability due to a pair of real multipliers when $f_b \in (8.45~\text{kHz}, 8.67~\text{kHz})$. As seen in Fig.~\ref{MI_actuator}(b), this instability is connected with the turning points of the energy of the DB as a function of its frequency (these occur when $dE/df_b=0$). Similar features have also been observed in diatomic Klein-Gordon chains~\cite{Gorbach2003}.

\begin{figure}[t]
\begin{center}
\includegraphics{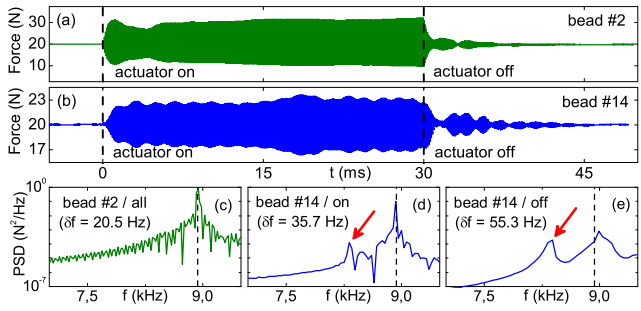}
\end{center} 
\caption{\label{expDB}[Color online] Experimental observation of a DB at $f_b^{\text{exp}} \simeq 8.35$~kHz. (a) Force at particle $i=2$, (b) force at particle $i=14$, (c) PSD at $i = 2$ for $t\geq1$~ms, (d) PSD at $i = 14$ for $1\leq t\leq29$~ms (while the actuator is on), and (e) PSD at $i = 14$ for $t\geq31$~ms (after the actuator is switched off). Vertical lines in (a,b) mark the times when the actuator is switched on and off. Vertical lines in (c,d,e) indicate the driving frequency ($8.94\mbox{~kHz} \gtrsim f_2^{\text{exp}}$; see Table~\ref{table_th_vs_exp_linear}), and $\delta f$ is the frequency resolution. Arrows in (d,e) indicate a DB that lies inside the band gap ($f_1^{\text{exp}}<f_b^{\text{exp}}<f_2^{\text{exp}}$; see Table~\ref{table_th_vs_exp_linear}).}
\end{figure}

{\it Experimental observation of DBs}. Here, we excite the 81-bead diatomic crystal with a higher-amplitude signal (relative to the linear-spectrum experiments). The actuator is driven by a $30$~ms long square windowed sine voltage. We probe a range of driving frequencies (near the lower optical cutoff frequency - see $f_2^{exp}$ in Table~\ref{table_th_vs_exp_linear}) and amplitudes around the values expected to create DBs. We place force sensors in particles $2$, $4$, $7$, $12$, and $14$. In Fig.~\ref{expDB}, we show experimental evidence of a DB. For the case shown in Fig.~\ref{expDB}, the driving frequency is $8.94\mbox{~kHz}$ and the peak force that we measure near the actuator is $12.20$~N~$\simeq$~0.61$F_0$. Figure~\ref{expDB}(b) shows the force versus time at particle $14$, and the corresponding PSDs, shown in Figs.~\ref{expDB}(d) and ~\ref{expDB}(e), demonstrate the existence of a second mode (whose frequency differs from the driving frequency), which we indicate with an arrow in the figure. This mode occurs at $f_b^{\text{exp}} \simeq 8.35$ kHz, which is in the band gap and yields a robust DB according to the previous linear stability analysis. As with the PSD at particle $2$, which we show in Fig.~\ref{expDB}(c), the PSDs at particles $4$ and $7$ do not exhibit additional modes in the gap. The PSD of the signal from sensor $12$ reveals the presence of this mode, but the amplitude is smaller than that from the sensor at $i=14$, indicating that the center of the DB is located further inside the bulk of the crystal. Additionally, as predicted by simulations, the DB appears to be long-lived, as shown for instance from the different decay rates in Figs.~\ref{expDB}(a) and~\ref{expDB}(b) after we switch off actuator. In Fig.~\ref{expDB}(e), we estimate the PSD of the tail, illustrating that the DB maintains its prominence while the mode at the actuator frequency has experienced a decrease in PSD amplitude by two orders of magnitude. 

{\it Conclusions}. We have characterized the dynamics of compressed 1D diatomic granular crystals using experiments, numerical simulations, and theoretical analysis. We found satisfactory agreement between experiments and theory for the linearized spectrum of steel-aluminum crystals. We also explored theoretically the formation of DBs via MI, which, in turn, led us to systematically trace them numerically and to observe them experimentally. Our results provide a first step towards achieving a deeper understanding and classifying intrinsic modes in 1D granular crystals, and pave the way for their manifestation and dynamics in 2D and 3D settings, leading ultimately towards their potential exploitation in energy-harvesting applications. 


{\it Acknowledgements}. We thank R. Carretero-Gonz\'{a}lez for useful discussions,
and NSF for support (Grants number: 0825983, 0844540, 0806762, 0349023).


\end{document}